# Scheduling Feasibility of Energy Management in Micro-grids Based on Significant Moment Analysis

Zhenwu Shi[1], Ningshi Yao[2], and Fumin Zhang[3]

Zhenwu Shi
C3 IoT, Redwood City, CA
E-mail: zhenwu.shi@c3iot.com

Ningshi Yao
School of Electrical and Computer Engineering, Georgia Institute of Technology, Atlanta, Georgia
E-mail: nyao6@gatech.edu

Fumin Zhang
Associate Professor
School of Electrical and Computer Engineering, Georgia Institute of Technology, Atlanta, Georgia
E-mail: fumin@gatech.edu

**Abstract** This paper studies the operation and scheduling of electric loads in micro-grid, a highly automated and distributed cyber-physical energy system (CPES). We establish rigorous mathematical expressions for electric loads and battery banks in the micro-grid by considering their characteristics and constraints. Based on these mathematical models, we propose a novel real-time scheduling analysis method for priority-based energy management in micro-grid, named Significant Moments Analysis (SMA). SMA pinpoints all the crucial moments when electrical operations are requested among the micro-grid and establishes a dynamic model to describe the scheduling behavior of electric loads. Using SMA, we can check the scheduling feasibility and predict whether the micro-grid can generate enough power to support the execution of electric loads. In the case where the power is insufficient to supply load demands, SMA can provide accurate information about the amount of insufficient power and the time when the insufficiency happens. Simulated results are presented to show the effectiveness of the proposed analysis method.

## 1 Introduction

Real-time scheduling is playing an important role in cyber-physical systems (CPS). Examples of cyber-physical systems range from small systems, such as medical equipment and automobiles, to large systems like national power grid. The majority of current CPSs require real-time managements on physical and cyber variables like battery state of charge, power flow, computing process, and network limitations. Because of the integration of physical and cyber world, cyber-physical systems must operate efficiently in real-time [Rajkumar et al., 2010]. The correctness of operations in CPSs depends not only on logical results of operations, but also on the time when these results are computed and transmitted to the physical plants. Therefore, real-time scheduling in cyber-physical systems is crucial for system stability and performance [Liu and Layland, 1973, Abdelzaher, Sharma and Chenyang Lu, 2004, Buttazzo et al., 2002, Lehoczky, 1990, Mok, 1983].

Cyber-physical energy system (CPES) studies the CPS in energy domain [Morris et al., 2009; Macana et al., 2011]. An important problem in CPES is the scheduling of the electric devices in real-time under power consumption/supply constraints. A properly scheduled CPES should balance the power usage well to avoid unfavorable conditions, such as peak energy consumption. Real-time scheduling of energy flow for CPES is a new topic that has emerged in recent years [Mohsenian-Rad et al., 2010; Palensky and Dietrich, 2011; Conejo et al., 2010; Mathieu et al., 2013]. The primary goal of real-time energy management is to schedule the operations of electric loads in electrical grids according to energy supply and user requirements [Subramanian et al., 2012]. However, real-time energy management has many new challenges that are different from scheduling in the traditional real-time operating system (RTOS). First, the available energy supply in CPES varies with respect to time, while the available computation resource in RTOS is fixed over the entire time period. Second, multiple electric loads may be scheduled at the same time in CPES, while RTOS devotes all resources to a single operation at any time instant. Finally, electric loads can be either preemptive or nonpreemptive depending on their functionalities.

Due to the above challenges, the scheduling analysis for RTOS cannot be directly applied to CPES. A number of pieces of work have been carried out on the scheduling analysis of CPES. Facchinetti et al. analyzed a type of electric load that controls the physical process [Facchinetti and Della Vedova, 2011]. The feasibility analysis in this work checks whether electric loads can be scheduled by real-time energy management so that specific constraints on the physical process are satisfied. Nghiem et al. studied the feasibility of scheduling electric loads under a given constrained peak power [Nghiem et al., 2011]. The results were further extended in Li et al. (2011), in which it was proved that the feasibility relies on the initial condition of electric loads.

In our work, we focus on the real-time energy management for a small and self-contained CPES called smart micro-grid [Lasseter and Paigi, 2014]. Micro-grid is a newly developed CPES that combines the advanced IT technology and can intelligently manage distributed power and energy storage devices [Huang et al., 2008]. One feature of micro-grid is that it utilizes renewable energy such as sunlight, wind, tides, and waves [Katiraei et al., 2008], which are more sustainable and environmental-friendly. With the integration of renewable energy, micro-grids are able to completely isolate themselves from the national electric grid, and function as a stand-alone grid to improve energy utilization efficiency and reliability. However, using renewable energy has challenges of providing consistent power supply because the renewable energy relies on the environmental power resources. Since the power supply in micro-grids is highly variable, it is even more challenging to schedule the electric loads in micro-grids. Plenty of research work has been done to study the energy management in micro-grids [Morais et al., 2010, Lee, Lai and Chan, 2011, Liu et al., 2015, Arboleya et al., 2015 , Barklund et al., 2008, Qin, Chen and Wang, 2012].

The feasibility analysis for micro-grids is primary in order to guarantee the independent operation. Operators of micro-grids need to check whether the electric loads can be scheduled under fluctuating renewable energy supply. In this chapter, we extend our previous work of the task scheduling for cyber-physical system [Zhang et al., 2008, 2009, 2013; Shi, 2014; Shi and Zhang, 2013; Wang et al., 2015; Zhang and Shi, 2009] and introduce a novel real-time scheduling analysis method, named Significant Moments Analysis (SMA). SMA serves as a centralized scheduling algorithm for coordinating discharge and charge of devices. Our contribution is as followings: 1) SMA analyzes scheduling behaviors of electric loads and gives out a sufficient and necessary condition for scheduling feasibility in micro-grids. 2) With the real-time sensing data and electric load demands, SMA determines the activation of each electric load based on its priority assignment. 3) SMA also provides accurate online predictions regarding when and how much power is insufficient for the independent operation of the micro-grid. Such information allows the operators of micro-grids to take necessary and preventive measures in advance. To the best of our knowledge, these contributions have not been documented in the literature.

This chapter is organized as follows. In section 2, we introduce the concept of micro-grid. In section 3, we present the mathematical definitions and expression for different components in the micro-grids. Based on the mathematical models, SMA is introduced in section 4. We use the state vector of SMA to describe the dynamic scheduling behavior in real-time energy management system. In section 5, the sufficient and necessary condition for scheduling feasibility is proposed based on SMA. Simulation results using SMA are presented in section 6. Section 7 is the conclusion and future work.

# 2 Smart Micro-grids

Micro-grids are modern, local, small-scale versions of CPES. With the integration of real-time data and control, micro-grids can operate independently from the national electric grid, which provides more security against terrorism and natural disasters [Kroposki et al., 2008]. Moreover, the use of renewable energy in micro-grids is environmental friendly. These benefits have greatly stimulated the adoption of micro-grids in various applications. For example, the military bases are actively deploying micro-grids in order to assure reliable power supply without relying on the national power system [Hayden, 2013]. Educational institutions also have extensively built micro-grids in their campuses. 90 percent of the annual electricity generation comes from micro-grids in University of California, San Diego. As discussed in [De Souza Ribeiro et al., 2010], micro-grids have become the best solution to isolated, stand-alone areas that may never be connected to the national electric grid due to their remoteness.

## 2.1 Infrastructure of micro-grids

We consider a typical micro-grid as shown in Figure 1, whose physical world consists of electric loads, on-site generations, and energy storage.

The left hand side in Figure 1 shows the on-site power supply, which comes from both the renewable energy [18] and fossil fuel generators. The power from on-site generations is clean, low cost, but highly variable. The right hand side in Figure 1 shows electric loads and battery banks. Batteries are energy storage devices, which can store energy whenever the supply exceeds the load demand and provide energy whenever the on-site generation are insufficient. The batteries provide a bridge to balance the supply and demand in micro-grids. The brown arrows represent the energy flow of micro-grids.

In micro-girds, all physical components are controlled by a centralized computer. The sensing data are transmitted to the computer. Based on the collected data, the computer calculates the control commands for battery bank and electric loads, and then sends the commands back to the loads. The dashed arrows represent the data flow in the cyber world of micro-grids.

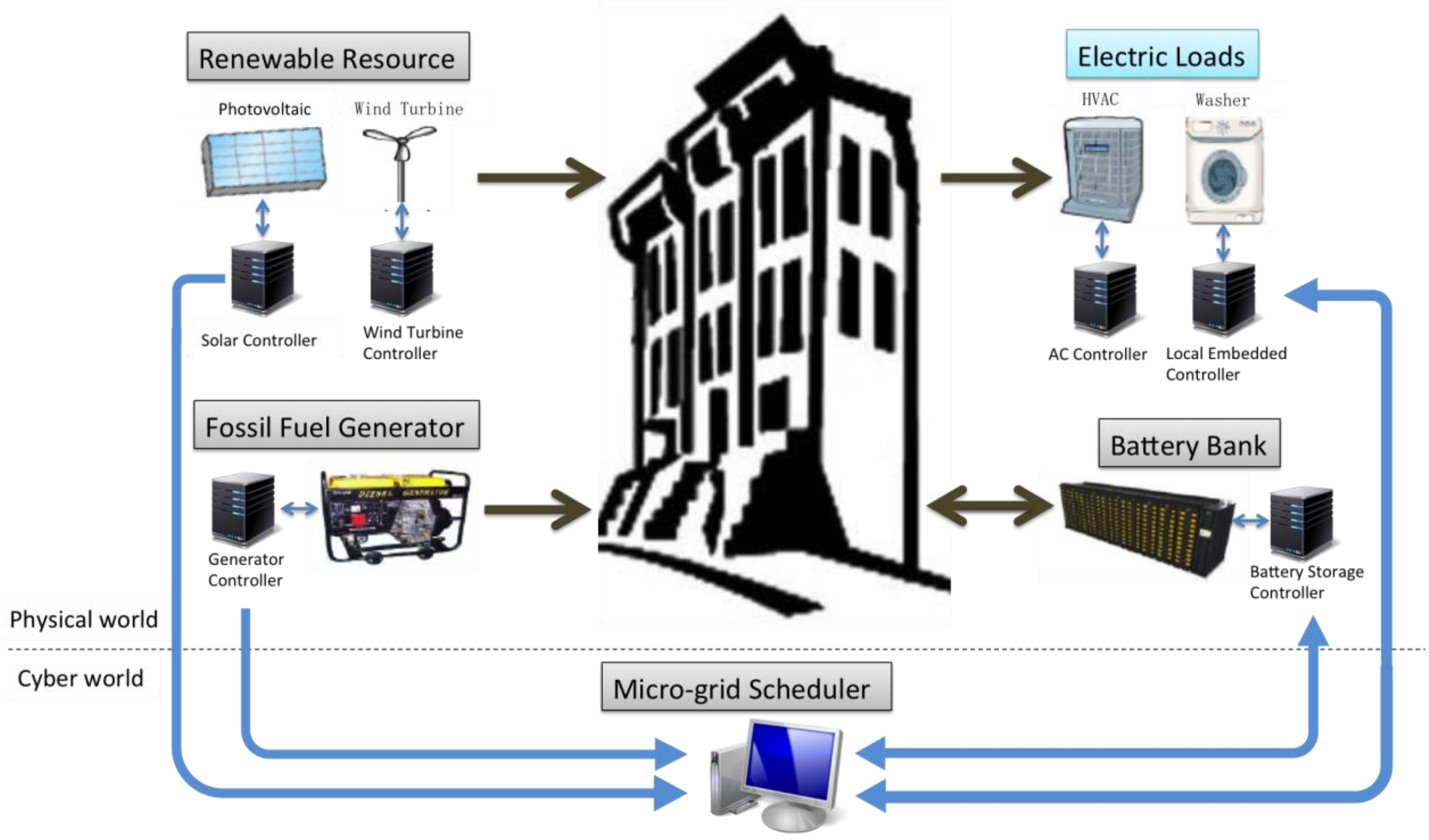

Fig. 1 An example of Micro grid

## 2.2 Independent Operation

A fully evolved micro-grid must have the ability to operate independently from the main electric grid for an extended period of time. Successful independent operation requires that the on-site generated power and the battery storage should meet the demand of electric loads. This requirement can be easily satisfied if all electric loads are deferrable.

However, in real applications, some electric loads in the micro-grid may become non-deferrable during the process of operation. We say an electrical load is non-deferrable if (1) the electrical load is non-preemptive and it is currently in the middle of execution; and (2) the electrical load cannot complete its execution before the deadline if not executed immediately. In this case, the total energy supply from the on-site generation and battery storage should be at least more than the total demand of all non-deferrable electric loads. However, identifying non-deferrable electric loads is not easy because an electrical load can switch between deferrable and non-deferrable state during its operation.

# 3 Models of micro-grids

In this section, we develop a set of models capable of describing different physical components in the micro-grids. Different from the previous real-time modeling of micro-gird [Subramanian et al., 2012, Facchinetti and Della Vedova, 2011], the models presented in this section capture the complicated functionalities of deferrable and non-deferrable electric loads. Based on these models, we can then analyze the dynamic behaviors of micro-grids in the next section.

## *3.1 Electric loads*

Without loss of generality, we assume that the micro-grid contains a set of electric loads $\Gamma = \{\tau_1, \cdots, \tau_N\}$. Each electrical load $t_n$ in G consists of a sequence of instances, and each instance corresponds to one operation request of $t_n$. We use $t_n[k]$ to denote the *k*-th instance of $t_n$. $t_n[k]$ can be characterized by the requested time $a_n[k]$, inter-request time interval $T_n[k]$ between $a_n[k]$ and $a_n[k+1]$, relative deadline $D_n[k]$, operational power $E_n[k]$, operation time $C_n[k]$, preemption $F_n[k]$, and priority $P_n[k]$. Note $F_n[k] = 0$ denotes that $t_n$ is non-preemptive during operation and $F_n[k] = 1$ denotes that $t_n$ is preemptive. The smaller value of $P_n[k]$ denotes a higher priority. In the following part of this section, we will study the representation of different types of electric loads using the above notations.

First, we give an example of a simple electrical load such as a rice cooker.

*Example 1.* A rice cooker operates once every 24 hours and each operation will take one hour. The operation is requested at 9:00am and must finish before 7:00pm. The operational power is 310*w*. The operation is non-preemptive once started. The rice cooker has the second highest priority.

Based on the above requirement, we can characterize the rice cooker with the following parameters:

$$\begin{aligned} &a_n[k] = 24k + 9, T_n[k] = 24, D_n[k] = 10, E_n[k] = 310, C_n[k] = 1 \\ &F_n[k] = 0, P_n[k] = 2 \end{aligned} \tag{1}$$

where $F_n[k]=0$ denotes that the electrical load is non-preemptive during its operation.

Second, we introduce electric loads with multiple internal operation phases. For example, the operation of dish washers goes through five phases as: pre-wash, wash, first rinse, drain, second rinse, and dry. The five operation phases must follow a strict sequential order. Each operation phase may have different execution time, power and preemption property. Table 1 shows a detailed specification of a dishwasher. According to the specification, we can express the dishwasher as:

$$E_n[k]=[64.20,\ 1517.8,\ 103.8,\ 8.2\ ,\ 1872.3,\ 10.9]$$
$$C_n[k]=[\ 0.25\ ,\ 0.54\ \ ,\ 0.17\ ,\ 0.07,\ \ 0.31\ ,\ 0.86]$$
$$F_n[k]=[\ \ 1\ ,\ \ 1\ \ ,\ \ 0\ ,\ 1\ ,\ \ 0\ ,\ \ 1\ ]$$

Table 1 Dishwasher Specification

| Phase | Pre-wash | Wash | 1st Rinse | Drain | 2nd Rinse | Dry |
|---|---|---|---|---|---|---|
| Operation Power | 64.20w | 1517.8w | 103.8w | 8.2w | 1872.3w | 1.9w |
| Operation Time | 0.25h | 0.54h | 0.17h | 0.07h | 0.31h | 0.86h |
| Preemption | Yes | Yes | No | Yes | No | Yes |

Third, we introduce electric loads subject to the precedence constraints with other electric loads. For instance, a dryer machine cannot start its operation until a washing machine has completed its operation. To model this, we can view any group of precedence constrained electrical load as a whole comprehensive electrical load, whose parameters $\{E_n[k], C_n[k], F_n[k], P_n[k]\}$ contains the characteristics of each electrical load. Therefore, the mathematical expression of comprehensive electric loads is similar to electric loads with internal operation phases (type 2). Note that $P_n[k]$ is a vector for comprehensive electric loads as each individual load has a different priority.

Finally, we introduce electric loads with dynamics changing according to the physical environment. Consider air conditioners (AC) as an example. We use $x$ to denote the house temperature inside the house, and $TP_{\text{out}}$ denote the outside temperature. According to the dynamic model in [Meliopoulos et al., 2013], we can represent the operation of AC as

$$\dot{x}=-\frac{G_{out}}{C_h}\left(x-TP_{out}\right)+\frac{1}{C_h}n_{ac}P_{ac}u \qquad (2)$$

where $G_{\text{out}}$ is the thermal conductance between house and outside environment, $C_h$ the thermal capacitance of the house, $n_{\text{ac}}$ the coefficient of performance of AC, $P_{\text{ac}}$ the power of AC, and $u$ is the duty cycle of AC. Note that AC will cycle on and off periodically. Therefore, the duty cycle of AC is the percentage of one period in which AC is on. By controlling the duty cycle $u$, AC will guarantee that the house temperature will stay within a bounded range such that $TP_{\min} \le x \le TP_{\max}$. Suppose that $x$ is currently at a stable point $x_{\text{stable}}$ such that $TP_{\min} \le x_{stable} \le TP_{\max}$. To guarantee that $x$ always stays at this point, we must have that $\dot{x} = 0$, i.e.

$$0 = -\frac{G_{out}}{C_h}\left(x - TP_{out}\right) + \frac{1}{C_h} n_{ac} P_{ac} u \tag{3}$$

which implies that

$$u = \frac{G_{out}}{n_{ac} P_{ac}}\left(x_{stable} - TP_{out}\right) \tag{4}$$

Given the duty cycle u, we have the execution time of AC as

$$C_n[k] = T_n[k]u = T_n[k]\frac{G_{out}}{n_{ac} P_{ac}}\left(x_{stable} - TP_{out}\right) \tag{5}$$

where $T_n[k]$ is the period of one on and off cycle. As it shows, the execution time of AC will dynamically change according to the outside temperature $TP_{\text{out}}$.

According to the above discussions, we can represent different types of electric loads as the tuple $\{C_n[k], E_n[k], D_n[k], T_n[k], F_n[k], P_n[k]\}$. Since we want to model the dynamics of the real-time energy management at any time $t$, we define the characteristics of electric loads in continuous time domain as follows

**Definition 1.** At any time $t$, an instance of $\tau_n$ is effective if and only if it is the nearest instance released before or at time $t$, i.e. $\tau_n[k]$ is effective at time $t$ if and only if

$$\alpha_n[k] \le t < \alpha_n[k+1] \tag{6}$$

**Definition 2.** At any time t, $C_n(t)$, $E_n(t)$, $D_n(t)$, $T_n(t)$, $F_n(t)$ and $P_n(t)$ are respectively defined as the operation time, power, relative deadline, inter-request time, preemption, and the priority of the effective instance of $\tau_n$, i.e.

$$\begin{gathered} \text{if } \alpha_n[k] \le t < \alpha_n[k+1] \\ C_n(t) = C_n[k],\ E_n(t) = E_n[k],\ D_n(t) = D_n[k],\ T_n(t) = T_n[k],\ F_n(t) = F_n[k],\ P_n(t) = P_n[k] \end{gathered} \tag{7}$$

Therefore, electric loads in the micro grid can be represented in continuous time domain as $\{C_n(t), E_n(t), D_n(t), T_n(t), F_n(t), P_n(t)\}_{n=1}^{N}$.

## *3.2 On-site Generation and Battery Bank*

In micro-grids, a noticeable portion of electricity is generated on-site from different sources of energy. We formally define on-site generation of electricity as follows

**Definition 3.** At any time *t*, EG(*t*) is defined as the on-site electricity generation in a micro-grid.

EG(*t*) includes the electricity from both the fossil fuel generator and the renewable energy. Figure 2 shows the power generation of one wind turbine from the Alberta Electric System Operator on July 12th, 2011. The maximum wind power generation at 10:00 is twice as much as the minimum wind power generation at 6:00. As it shows, the electricity generated from renewable energy is highly variable, so the total on-site electricity generation EG(*t*) changes with respect to time.

Battery bank is used in micro-grid applications where the generation and load demand cannot be exactly matched. It increases the stability and reliability of the micro-grids. We formally define the battery bank as follows

**Definition 4.** At any time *t*, SOC(*t*) is defined as the state of charge of the battery bank in the micro-grid. $B_{power}$ is defined as maximum charge/discharge rate of the battery bank. $B_{capacity}$ is defined as capacity of the battery bank.

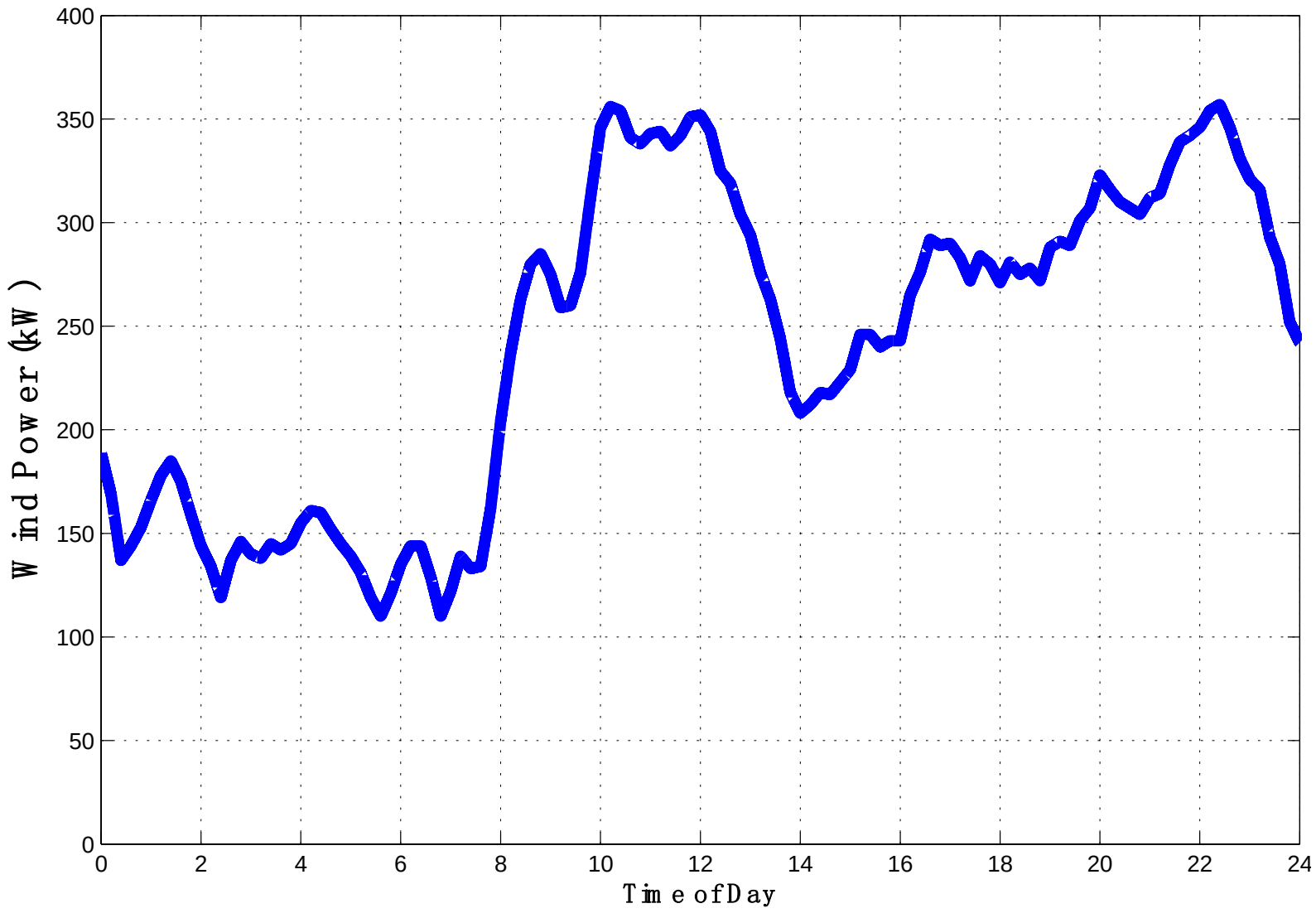

Fig. 2 Wind Power Generation reported by Alberta Electric System Operator

SOC(*t*) indicates the percentage of energy remaining in the battery bank. In real applications, batteries should not be discharged below 20% of its SOC. Otherwise, the battery life will be significantly shortened. Therefore, we put the following constraints on the battery operation such that

$$20\% \le SOC(t) \le 100\% \qquad (8)$$

According to the above constraint, we know that the maximum power output of the battery bank at any time *t* depends on its state of charge. When the state of charge is larger than 20%, we have the maximum output power of the battery bank as $B_{power}$, i.e.

when $SOC(t)$ > 20%, maximum power output of the battery bank is $B_{power}$ (9)

On the other hand, when the state of charge drops to 20%, the batteries is not allowed to output any power, i.e.

$$\text{when } SOC(t) £ 20\%, \text{ maximum power output of the battery bank is 0} \qquad (10)$$

According to Equation (9) and (10), the maximum power output of the battery bank at any time $t$ is expressed as

$$\text{sgn}(SOC(t) - 20\%)B_{power} \qquad (11)$$

# 4 Significant Moment Analysis

In the previous sections, we have shown that the micro-grid consists of three major components represented as follows

1. On-site electricity generation is represented as EG($t$);

2. Batteries are represented as {SOC($t$), $B_{power}$, $B_{capacity}$};

3. Electric loads $\{C_n(t), E_n(t), D_n(t), T_n(t), F_n(t), P_n(t)\}_{n=1}^{N}$, where $N$ denotes the total number of electric loads in the micro-grid.

The establishment of these mathematical expressions will allow us to analytically study the dynamics of real-time energy management in the micro-grid. Compared to the critical time instant analysis introduced by Liu and Layland in [2], which is a traditional real-time scheduling analysis studying the worst-case scenarios, the analysis method proposed in this chapter studies all the significant moments whenever a scheduling decision is required. This analysis method is referred as Significant Moment Analysis (SMA).

## *4.1 State Vector of Electric loads*

We consider a micro-grid with a set of electricity loads $\{\tau_1,\cdots,\tau_N\}$. To study the progression of electric loads under the real-time energy management, we introduce a state vector $Z(t) = [S(t), R(t), O(t)]^T$ for SMA that describes the current status of $\{\tau_1,\cdots,\tau_N\}$ at any time $t$.

**Definition 5.** The dynamic inter-request time is defined as $S(t) = [s_1(t),\cdots,s_N(t)]$, where $s_n(t)$, for $n$ = 1, 2, ⋯, $N$, denotes how long after $t$ the next instance of $t_n$ will be requested.

**Definition 6.** The remaining time is defined as $R(t) = [r_1(t),\cdots,r_N(t)]$, where $r_n(t)$, for $n$ = 1, 2, ⋯, $N$, denotes the remaining operation time of the current instance of $t_n$ *after* time $t$.

**Definition 7.** The dynamic response time is defined as $O(t) = \left[o_1(t), \cdots, o_N(t)\right]$, where $o_n(t)$, for $n = 1, 2, \cdots, N$, denotes how much time has elapsed before the current instance of $t_n$ finishes operation.

Based on the above definitions, we know that the progression of electric loads under the real-time energy management can be represented through the evolution of the SMA state vector $Z(t)$. In the following part, we will study the evolution of $Z(t) = \left[S(t), R(t), O(t)\right]^T$ under real-time energy management.

## *4.2 Non-Deferrable Electric loads*

According to Section 2.2, two types of electric loads are considered as non- deferrable. In this subsection, we can easily identify all the non-deferrable loads using the state vector $Z(t)$ of SMA.

The first type of non-deferrable electric loads could not complete execution before deadline if further delayed. This type of electrical load is represented as

$$\left\{ i \mid o_i(t) + r_i(t) = D_i(t) \right\} \qquad (12)$$

where $o_i(t)$ denotes the delay of $t_i$ since its time of request, $r_i(t)$ denotes the remaining operation of $t_i$, and $o_i(t)+r_i(t) = D_i(t)$ indicates that $t_i$ can complete before its deadline if and only if it continuously executes from now on.

The second type of non-deferrable electric loads are electric loads that are non-preemptive and currently in the middle of operation. This type of electric loads is represented as

$$\left\{ i \mid F_i(t) = 0, \text{ and } 0 < r_i(t) < C_i(t) \right\} \qquad (13)$$

where $F_i(t) = 0$ denotes that $t_i$ is non-preemptive, and $0 < r_i(t) < C_i(t)$ denotes that $t_i$ is in the middle of operation.

**Definition 8.** We use NonDefer($t$) to denote a set of non-deferrable electric loads that must be executed immediately at time $t$. NonDefer($t$) can be represented as

$$\text{Nondefer}(t) = \left\{ i \mid o_i(t) + r_i(t) = D_i(t) \right\} \cup \left\{ i \mid F_i(t) = 0, \text{ and } 0 < r_i(t) < C_i(t) \right\} \qquad (14)$$

## *4.3 Real-time Energy Management*

In this section, we rigorously define real-time energy management using mathematical expression. Let $\{1, \cdots, N\}$ denote the set of indices of total electric loads in the micro-grid and OP($t$) denote the set of indices of electric loads executing at time $t$.

**Definition 9.** A *real-time energy management* is a set-valued map between $R^+$ and the collection of all subsets of $\{1, \cdots, N\}$. It is parameterized as OP($t$) : $R^+ \rightarrow 2^{\{1,\cdots,N\}}$, where $t \in R^+$.

At any time $t$, OP($t$) should at least contain non-deferrable electric loads. If the energy supply is larger than the demand of all non-deferrable electric loads, the real-time energy management will activate the execution of other deferrable electric loads in terms of their priorities.

At any time t, we can construct the real-time energy management OP($t$) through three steps as follows

**First Step: Initialization.** Initialize OP($t$) as a set of indices of non-deferrable electric loads at time $t$, i.e.

$$\mathrm{OP}(t) = \mathrm{Nondefer}(t) \tag{15}$$

Moreover, we initialize an scheduling pool SCH($t$) as a set of deferrable electric loads that have remaining operation time, i.e.

$$\mathrm{SCH}(t) = \left\{ i \mid r_i(t) > 0 \right\} - \mathrm{Nondefer}(t) \tag{16}$$

where $\left\{ i \mid r_i(t) > 0 \right\}$ denotes a set of electric loads with remaining operation time and "-" denotes the substraction between two sets.

**Second Step: Scheduling.** In this step, we decide whether the highest priority electrical load in SCH($t$) can be scheduled to operate at time $t$. The highest priority electrical load in SCH($t$) can be denoted as $\tau_n$ such that $n = \min_{i \in \mathrm{SCH}(t)} P_i(t)$. If $\tau_n$ satisfies the following condition,

$$E_n(t) + \sum_{i \in \mathrm{OP}(t)} E_i(t) \le EG(t) + \mathrm{sgn}\left(SOC(t) - 20\%\right) B_{power} \tag{17}$$

where $\sum_{i \in \mathrm{OP}(t)} E_i(t)$ represents the demand of electric loads that have been scheduled to execute at time $t$, $E_n(t)$ represents the demand of $\tau_n$, and the right hand side represents the total supply of on-site generation and the battery storage. Equation (17) indicates that the total supply at current time $t$ is enough to satisfy the demand of both electric loads in OP($t$) and the electrical load $\tau_n$. In this case, $\tau_n$ will be scheduled to execute at time $t$, i.e.

$$\mathrm{OP}(t) = \mathrm{OP}(t) + \{n\} \tag{18}$$

On the other hand, if the electrical load $\tau_n$ satisfies the following condition,

$$E_n(t) + \sum_{i \in \mathrm{OP}(t)} E_i(t) > EG(t) + \mathrm{sgn}\left(SOC(t) - 20\%\right) B_{power} \tag{19}$$

which indicates that the total energy at current time $t$ is NOT enough to satisfy the demand of both electric loads in OP($t$) and the electrical load $t_n$. In this case, $t_n$ will NOT be scheduled to execute at time $t$, i.e.

$$\mathrm{OP}(t) = \mathrm{OP}(t) \qquad (20)$$

**Third Step: Update and Check.** In this step, we update scheduling pool SCH($t$) and check the number of electric loads in SCH($t$). Since the execution of electrical load $t_n$ has been decided in the second step, we need to remove $t_n$ from the scheduling pool, i.e.

$$\mathrm{SCH}(t) = \mathrm{SCH}(t) - \{n\} \qquad (21)$$

If the scheduling pool is NOT empty, i.e. $\mathrm{SCH}(t) \neq \emptyset$, real-time energy management needs to decide the execution of other electric loads in SCH($t$). In this case, go back to Step 2. On the other hand, if the scheduling pool is empty, i.e. $\mathrm{SCH}(t) = \emptyset$, the execution of all deferrable electric loads have been decided and the construction of OP($t$) finishes.

Algorithm 1 gives a detailed implementation of real-time energy management. The input of the algorithm is the current SMA state vector $Z(t)$, on-site generation, and battery SOC. The output of the algorithm is a set of electric loads that are scheduled to execute at time $t$. In the next section, we will show that at any time $t$, an electrical load $t_n$ has different evolution dynamics in SMA, depending on whether $t_n$ is scheduled to execute (i.e. $n \in \mathrm{OP}(t)$) or not (i.e. $n \notin \mathrm{OP}(t)$).

**Algorithm 1:** Real-time Energy Management

**Data**: $Z(t)$, $\mathrm{SOC}(t)$, $\mathrm{EG}(t)$, $\mathscr{B}_{\mathrm{power}}$, $\{C_n(t), E_n(t), D_n(t), T_n(t), F_n(t), P_n(t)\}_{n=1}^{N}$
**Result**: OP($t$)

$\mathrm{NonDefer}(t) = \{i \mid o_i(t) + r_i(t) = D_i(t)\} \cup \{i \mid F_i(t) = 0, \text{and } 0 < r_i(t) < C_i(t)\}$;

/*1st Step: Initialization*/

$\mathrm{OP}(t) = \mathrm{NonDefer}(t)$;

$\mathrm{SCH}(t) = \{i \mid r_i(t) > 0\} - \mathrm{NonDefer}(t)$;

**while** $\mathrm{SCH}(t) \neq \emptyset$ **do**

  /*2nd Step: Scheduling*/

  $n = \max_{i \in \mathrm{SCH}(t)} P_i(t)$ ;

  **if** $\sum_{i \in \mathrm{OP}(t)} E_i(t) + E_n(t) \leq \mathrm{EG}(t) + \mathrm{sgn}(\mathrm{SOC}(t) - 20\%)\mathscr{B}_{\mathrm{power}}$ **then**

    $\mathrm{OP}(t) = \mathrm{OP}(t) + \{n\}$;

  **else**

    $\mathrm{OP}(t) = \mathrm{OP}(t)$;

  /*3nd Step: Update and Check*/

  $\mathrm{SCH}(t) = \mathrm{SCH}(t) - \{n\}$;

**return** OP($t$);

## *4.4 Evolution of Electric loads in SMA*

The SMA state vector $Z(t)$ contains the status for a set of $N$ electric loads. Therefore, the evolution of $Z(t)$ can be obtained through the evolution of each electrical load. For each electrical load $\tau_n$, its state vector $\left[s_n(t),\ r_n(t),\ o_n(t)\right]^T$ can evolve in both continuous and discrete ways.

First, we discuss the discrete evolution of $\left[s_n(t),\ r_n(t),\ o_n(t)\right]^T$. If a new instance of $\tau_n$ arrives at time $t$, i.e. $s_n(t) = 0$, the state vector will update according to the characteristics of the new instance. Thus, we have that:

$$\text{if } s_n(t)=0: \begin{bmatrix} s_n(t) \\ r_n(t) \\ o_n(t) \end{bmatrix} = \begin{bmatrix} T_n(t) \\ C_n(t) \\ 0 \end{bmatrix} \tag{22}$$

Next, we discuss the continuous evolution of $\left[s_n(t),\ r_n(t),\ o_n(t)\right]^T$. According to Section 4.3, the execution of $\tau_n$ will continue at time $t$ when $n \in \text{OP}(t)$ and stop at time $t$ when $n \notin \text{OP}(t)$. Therefore, the continuous evolution of $\left[s_n(t),\ r_n(t),\ o_n(t)\right]^T$ depends on the execution status of $\tau_n$.

**Case 1:** $\tau_n$ will execute at time $t$, i.e. $n \in \text{OP}(t)$. Then the continuous evolution of $\left[s_n(t),\ r_n(t),\ o_n(t)\right]^T$ at time $t$ is

$$\text{if } s_n(t) \neq 0 \text{ and } n \in \text{OP}(t): \begin{bmatrix} \dot{s}_n(t) \\ \dot{r}_n(t) \\ \dot{o}_n(t) \end{bmatrix} = \begin{bmatrix} -1 \\ -1 \\ 1 \end{bmatrix} \tag{23}$$

**Case 2:** $\tau_n$ will NOT execute at time $t$, i.e. $n \notin \text{OP}(t)$. In this case, $s_n(t)$ will still continuously decrease. And since the execution of $\tau_n$ stops at time $t$, the remaining time $r_n(t)$ will NOT decrease in this case because, i.e. $r_n(t+\Delta t) = r_n(t)$. Finally, the evolution of $o_n(t)$ from $t+\Delta t$ is a little involved. It depends whether the current instance of $\tau_n$ has completed its execution before time $t$. If the current instance of $\tau_n$ has completed execution before time $t$, i.e. $r_n(t) = 0$, the dynamic response time $o_n(t)$ keeps constant. On the other hand, if the current instance of $\tau_n$ has NOT completed execution before time $t$, i.e. $r_n(t) > 0$, the dynamic response time $o_n(t)$ increases, i.e. $r_n(t+\Delta t) = r_n(t)$. In all, the continuous evolution of $\left[s_n(t),\ r_n(t),\ o_n(t)\right]^T$ at time $t$ can be expressed as

$$\text{if } s_n(t) \neq 0 \text{ and } n \notin \text{OP}(t): \begin{bmatrix} \dot{s}_n(t) \\ \dot{r}_n(t) \\ \dot{o}_n(t) \end{bmatrix} = \begin{bmatrix} -1 \\ 0 \\ \text{sgn}\left(r_n(t)\right) \end{bmatrix} \quad (24)$$

## *4.5 Battery State of Charge Evolution*

As discussed before, the batteries in the micro-grid have two modes of operation. In the first mode, the on-site electricity generation exceeds the load demand and the batteries store extra energy. This happens when the following condition is satisfied $\sum_{i\in \text{OP}(t)} E_i(t) < EG(t)$. The extra power stored in batteries at time $t$ is $EG(t) - \sum_{i\in \text{OP}(t)} E_i(t)$. In the second mode, the on-site energy is insufficient to supply load demands and the batteries will provide power to electric loads. It happens when the following condition is satisfied $\sum_{i\in \text{OP}(t)} E_i(t) > EG(t)$. The power provided by batteries is $\sum_{i\in \text{OP}(t)} E_i(t) - EG(t)$.

We use $\Delta t$ to denote an arbitrarily small time. Then, for both two modes, we have that

$$\left(SOC(t+\Delta t) - SOC(t)\right) B_{capacity} = \left(EG(t) - \sum_{i\in \text{OP}(t)} E_i(t)\right) \Delta t \quad (25)$$

which implies that

$$\dot{SOC} = \lim_{\Delta t \to 0} \frac{SOC(t+\Delta t) - SOC(t)}{\Delta t} = \frac{EG(t) - \sum_{i\in \text{OP}(t)} E_i(t)}{B_{capacity}} \quad (26)$$

# 5 Feasibility Analysis

In this section, we will propose a necessary and sufficient feasibility analysis enable by SMA that checks whether the requirement for the independent operation of the micro-grid can be satisfied. Based on this feasibility analysis, we can also accurately predict when and how much power is insufficient for the independent operation of micro-grids.

## *5.1 Necessary and Sufficient Feasibility Analysis*

Our feasibility analysis of real-time energy management checks whether the micro-grids can operate independently from the main electric grid. Here, we propose a necessary and sufficient condition for feasibility analysis as follows

*Claim.* A micro-grid can operate independently from the main electric grid within $[t_a, t_b]$ if and only if it satisfies the following conditions for all $t \in [t_a, t_b]$,

$$\sum_{i\in \text{NonDefer}(t)} E_i(t) \le EG(t) + \text{sgn}\left(SOC(t) - 20\%\right) B_{power} \quad (27)$$

*Proof.* According to Definition 8 and Equation (14), NonDefer($t$) denotes a set of non-deferrable electric loads that must be executed at time $t$, which can be identified by SMA. Since $E_i(t)$ denotes the power demand of one non-deferrable electrical load $\tau_i$, the left hand side of Equation (27) denotes the demand of all non-deferrable electric loads at current time $t$.

On the other hand, EG($t$) denotes the on-site generation according to Definition 3, and $\mathrm{sgn}\left(SOC(t) - 20\%\right) B_{power}$ denotes the maximum power output of the battery storage. Hence, the right hand side of Equation (27) denotes the total supply from both the on-site generation and battery storage at time $t$.

Therefore, the micro-grid can run independently at time $t$ if and only if the left hand side (power demand) is smaller or equal to the right hand side (power supply). The micro-grid can run independently within $\left[t_a, t_b\right]$ if and only if the inequality condition is satisfied for any time $t \in \left[t_a, t_b\right]$.

*Note.* The *Claim* seems to be quite straightforward. But how to check whether the condition (27) is satisfied for all time $t$ is a very difficult task. The main difficulties are: first, the set of NonDefer(t) can not be determined unless we know the SMA states, and second the batter state SOC(t) needs to be predicted. The SMA equations (22), (23) and (24), together with the SOC equation (26) and Algorithm 1 are used to determine NonDefer(t) and SOC(t) so that condition (27) can be checked. This is the main contribution of our work.

## *5.2 Deficiency in Power Supply*

A successful feasibility analysis in Claim 5.1 checks whether the micro-grid can run independently from the main electric grid. Based on this analysis, we can predict when and how much power is insufficient for the independent operation of the micro-grids.

The power deficiency depends on the relation between the power demand and supply.

**Case1:** the power demand of all non-deferrable electric loads is smaller or equal to the power supply from the on-site generation and the battery storage, i.e.

$$\sum_{i \in \mathrm{NonDefer}(t)} E_i(t) \le EG(t) + \mathrm{sgn}\left(SOC(t) - 20\%\right) B_{power} \qquad (28)$$

In this case, the micro-grid can run independently from the main electric grid.

**Case2:** the power demand of all non-deferrable electric loads is larger than the power supply from the on-site generation and the battery storage, i.e.

$$\sum_{i \in \mathrm{NonDefer}(t)} E_i(t) > EG(t) + \mathrm{sgn}\left(SOC(t) - 20\%\right) B_{power} \qquad (29)$$

In this case, the micro-grid cannot run independently and the amount of insufficient power supply can be expressed as

$$\sum_{i \in \text{NonDefer}(t)} E_i(t) - EG(t) - \text{sgn}\left(SOC(t) - 20\%\right) B_{power} \qquad (30)$$

Based on the above analysis, we have the following claim about the insufficient power supply of the micro-grid.

*Claim.* At any time *t*, the amount of insufficient power for the independent operation of micro-grid can be expressed as

$$\max\left\{0, \sum_{i \in \text{NonDefer}(t)} E_i(t) - EG(t) - \text{sgn}\left(SOC(t) - 20\%\right) B_{power}\right\} \qquad (31)$$

# 6 Numeric Simulation

In this section, we will use numeric simulations to verify our feasibility analysis of real-time energy management in the micro-grid. All simulation results are implemented in MATLAB.

## *6.1 Simulation Setup*

First, we consider a micro-grid with the on-site renewable energy and fossil fuel generation. The renewable energy has the generation as shown in Figure 2 and the fossil fuel can output the constant power of 100kW.

Then, we consider four types of electric loads in the micro-grid. The first type of electric loads has simple characteristics as follows:

$$\begin{aligned}
&C_1(t) = 0.5\; E_1(t) = 80 \quad D_1(t) = 2\; T_1(t) = 2\; F_1(t) = 0\; P_1(t) = 1 \\
&C_2(t) = 0.5\; E_2(t) = 120\; D_2(t) = 2\; T_2(t) = 3\; F_2(t) = 1\; P_2(t) = 2 \\
&C_3(t) = 1.0\; E_2(t) = 160\; D_2(t) = 4\; T_2(t) = 4\; F_2(t) = 1\; P_2(t) = 3
\end{aligned} \qquad (32)$$

The second type of electric loads has multiple internal operation phases and can be represented as

$$\begin{aligned}
&C_4(t) = [0.5,\ 1] \quad E_4(t) = [150,\ 160] \quad D_4(t) = 4\; T_4(t) = 4\; F_4(t) = [0,\ 0] \quad P_4(t) = 4 \\
&C_5(t) = [1,\ 0.5,\ 1]\; E_5(t) = [120,\ 80,\ 180]\; D_5(t) = 4\; T_5(t) = 5\; F_5(t) = [0,\ 0,\ 0]\; P_5(t) = 5
\end{aligned} \qquad (33)$$

The third type of electric loads has precedence constraints and they formulate an comprehensive electric loads that can be represented as

$$\begin{aligned}
&C_6(t) = [0.5,\ 1,\ 0.5,\ 1.5,\ 0.5]\; E_6(t) = [50,\ 120,\ 30,\ 140,\ 80] \\
&D_6(t) = 6\; T_6(t) = 6\; F_5(t) = [0,\ 1,\ 0,\ 1,\ 0]\; P_6(t) = [6,\ 7,\ 8,\ 9,\ 10]
\end{aligned} \qquad (34)$$

The last type of electrical load is an AC operating dynamically according to the outside temperature $TP_{out}$. According to Equation (5), we have this electrical load represented

$$C_7(t) = T_7(t)u \; E_7(t) = 120 \; D_7(t) = 2 \; T_7(t) = 2 \; F_7(t) = 0 \; P_7(t) = 11 \qquad (35)$$

where $u$ is the duty cycle as $u = \frac{1}{400}\left(70 - TP_{out}\right)$

Finally, we assume the battery bank has the nominal voltage of 400V and the capacity of 450Ah. The C-rate of the battery bank is 0.5. The value of $B_{capacity}$ and $B_{power}$ are as following:

$$B_{capacity} = 180\text{kWh} \quad B_{power} = 90\text{kWh} \qquad (36)$$

Figure 3 shows the power demand of the above electric loads in the micro-grid under the real-time energy management. The solid blue line represents the total on-site energy generation including the wind power and fossil fuel. The area with cross line represents the power demand of all electric loads within 24 hours. As it shows, the power demand exceeds the on-site generation at some time points. At these time points, if the battery can provide enough energy to cover the extra power demand, the micro-grid will still be able to run independently. Otherwise, the micro-grid cannot run independently.

## *6.2 Feasibility Verification*

In this section, we will show that our feasibility analysis can accurately predict when and how much power is insufficient for the independent operation of the micro-grids.

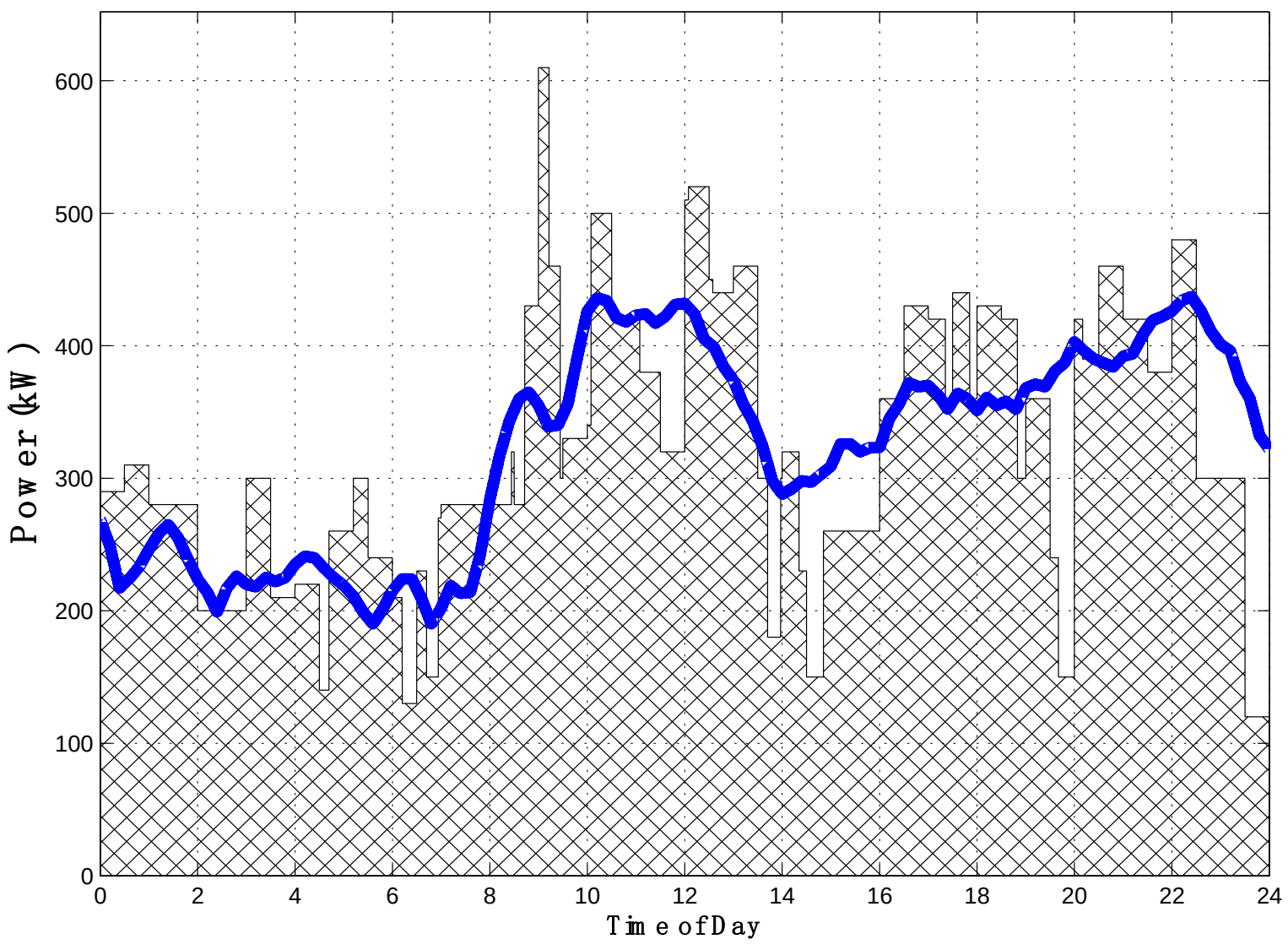

Fig. 3 Power Demand under Real-time Energy Management

As shown in Figure 3, the power demand exceeds the on-site generation at some time points. To guarantee the independent operation of the micro-grid, the battery storage will provide extra energy to electric loads. Figure 4 shows the battery dynamics within 24 hours. The green line represents the SOC of the battery. The evolution of the battery SOC is derived as part of the dynamic timing model in Section 3.2. The battery SOC increases when the battery is charging and decreases when the battery is discharging. For example, we can see from Figure 3 that the battery is charging within the time interval [14.4, 16] because the on-site generation is larger than the total power demand. This observation exactly matches the increase of the battery SOC within [14.4, 16] as shown in Figure 4. Moreover, the solid red area represents the discharging power of the battery. The faster the battery SOC decreases, the larger the discharging power is.

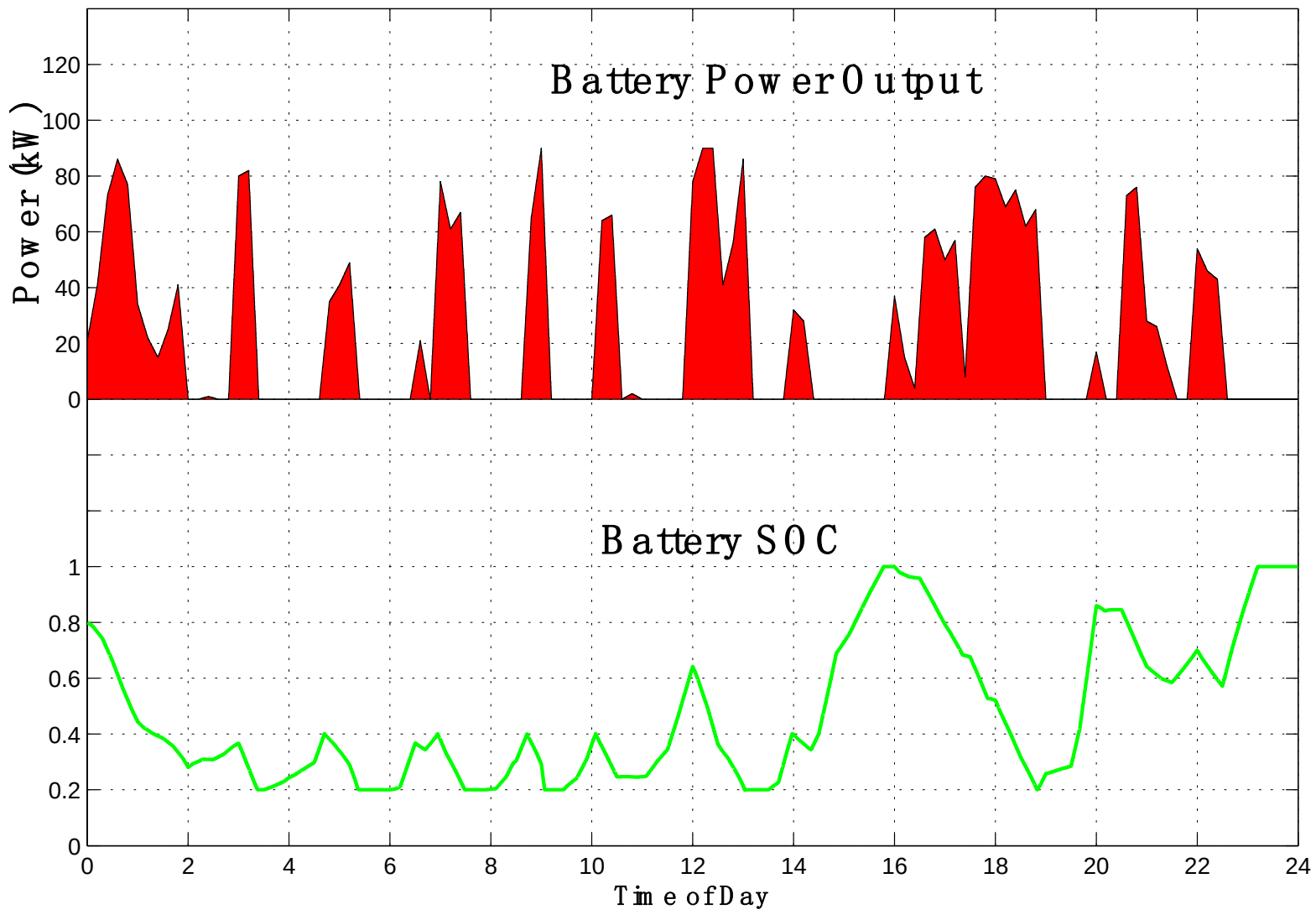

Fig. 4 Battery output and SOC

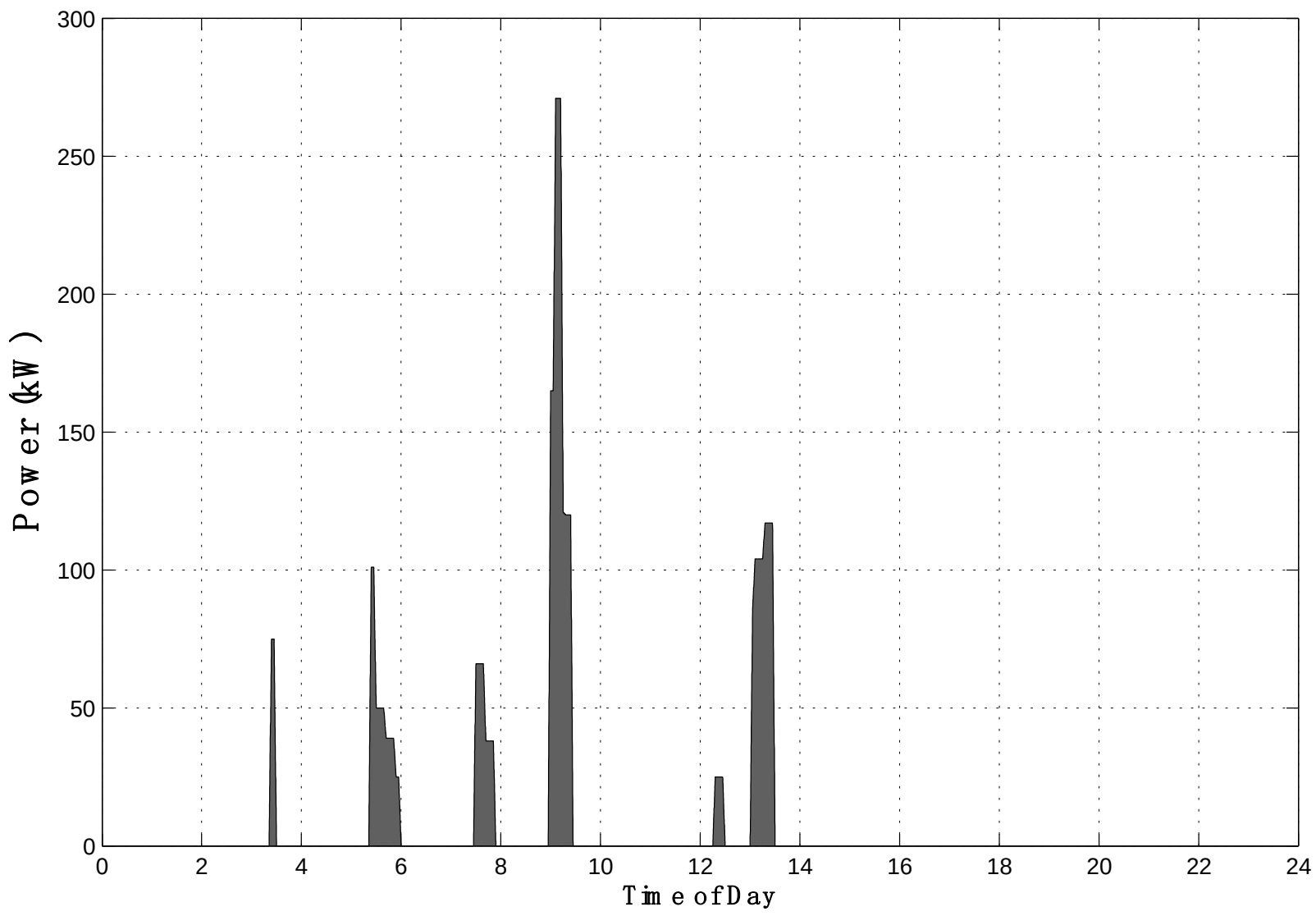

Fig. 5 Power Deficiency

However, even with the supply from the battery storage, the micro-grid may still not be able to run independently. Figure 5 shows when and how much power is insufficient for the independent operation of micro-grids. It is derived from our feasibility analysis studied in Section 5.

To verify the result of the feasibility analysis, we mark the results of Figure 4 and Figure 5 using different colors, as illustrated in Figure 6. The red color denotes the discharging power of the battery, which is same as that in Figure 4. The grey color denotes the amount of insufficient power for the independent operation of the micro-grids, which is same as that in Figure 5. As we can see, the combination of the red area and grey area exactly cover the extra power demand exceeding the on-site energy generation (solid blue line). Therefore, our feasibility analysis can accurately predict when and how much power is insufficient for the independent operation of the micro-grid.

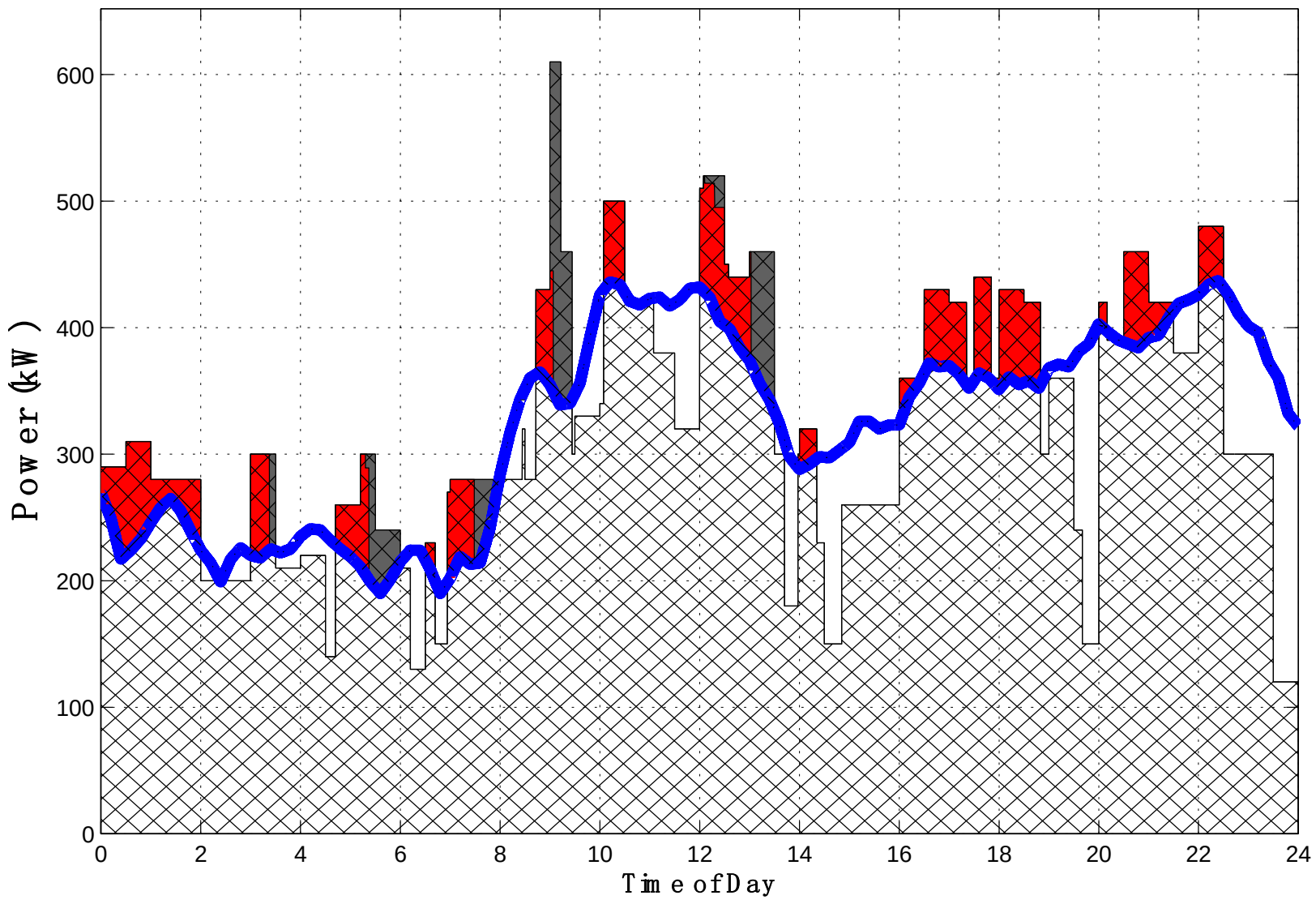

Fig. 6 Power Demand under Real-time Energy Management

# 7 Conclusions and Future Work

The main contribution of this work is the novel Significant Moment Analysis (SMA) for real-time energy management in CPES. We introduce SMA state vectors to describe dynamic behaviors of multiple operations scheduled in smart micro-grids. Based on SMA, we can easily find out all the non-deferrable electric loads and schedule the deferrable loads according to the current power supply at any moment.

Feasibility analysis is a fundamental step that guarantees the normal operation of the micro-grid under appropriate energy management. One of the most important motivations behind the micro-grids is the economic benefit. The future research direction of energy management can focus on optimizing the operation or implementation cost of the micro-grid, under the feasibility constraint. An- other future direction can be the feasibility analysis of a cluster of micro-grids. Each micro-grid is not running independently but connected with other micro-grids. Such connection among micro-grids will improve the robustness and the stability of the whole system.

**Acknowledgements** This work is partially supported by ONR grants N00014-10-10712 (YIP) and N00014-14-1-0635; and NSF grants OCE-1032285, IIS-1319874, and CMMI-1436284. The authors want to thank for the support.